\documentclass[10pt,prx,aps,notitlepage,superscriptaddress,onecolumn]{revtex4-1}

\usepackage[utf8]{inputenc}
\usepackage[T1]{fontenc}
\usepackage[english]{babel}  

\usepackage{tkz-graph}
\usepackage{tikz}

\usepackage{colortbl}
\usepackage{tabu,diagbox}

\usepackage{times}
\usepackage{dsfont}

\usepackage{amssymb,amsmath,amsfonts,amsthm}
\usepackage{mathtools}
\usepackage{verbatim}
\usepackage{enumerate}
\usepackage{graphicx}
\graphicspath{figs/}
\usepackage[colorlinks,linkcolor=red,anchorcolor=blue,urlcolor=blue,citecolor=blue]{hyperref}
\usepackage[capitalise]{cleveref}
\usepackage{bbm}
\usepackage{booktabs}

\usepackage{subfig}
\usepackage{caption}

\usepackage{color}
\definecolor{dred}{rgb}{.8,0.2,.2}
\definecolor{ddred}{rgb}{.8,0.5,.5}
\definecolor{dblue}{rgb}{.2,0.2,.8}
\definecolor{dgreen}{rgb}{.2,0.5,.2}

\newcommand{\bra}[1]{\mbox{$\langle #1|$}}
\newcommand{\ket}[1]{\ensuremath{|#1\rangle}}

\DeclareMathOperator{\Tr}{Tr}

\newcommand{\be}{\begin{equation}}
\newcommand{\ee}{\end{equation}}
\newcommand{\bea}{\begin{eqnarray}}
\newcommand{\eea}{\end{eqnarray}}

\begin{document}

\title{Local-measurement-based quantum state tomography via neural networks}

\author{Tao Xin}
\thanks{These authors contributed equally to this work.}
\affiliation{Shenzhen Institute for Quantum Science and Engineering and Department of Physics, Southern University of Science and Technology, Shenzhen 518055, China}
\affiliation{State Key Laboratory of Low-Dimensional Quantum Physics and Department of Physics, Tsinghua University, Beijing 100084, China}

\author{Sirui Lu}
\thanks{These authors contributed equally to this work.}
\affiliation{State Key Laboratory of Low-Dimensional Quantum Physics and Department of Physics, Tsinghua University, Beijing 100084, China}

\author{Ningping Cao}
\thanks{These authors contributed equally to this work.}
\affiliation{Department of Mathematics \& Statistics, University of Guelph, Guelph N1G 2W1, Ontario, Canada}%
\affiliation{Institute for Quantum Computing, University of Waterloo, Waterloo N2L 3G1, Ontario, Canada}

\author{Galit Anikeeva}
\affiliation{Institute for Quantum Computing,
University of Waterloo, Waterloo N2L 3G1, Ontario, Canada}

\author{Dawei Lu}
\affiliation{Shenzhen Institute for Quantum Science and Engineering and Department of Physics, Southern University of Science and Technology, Shenzhen 518055, China}

\author{Jun Li}
\email{lij3@sustc.edu.cn}
\affiliation{Shenzhen Institute for Quantum Science and Engineering and Department of Physics, Southern University of Science and Technology, Shenzhen 518055, China}
\affiliation{Institute for Quantum Computing,
University of Waterloo, Waterloo N2L 3G1, Ontario, Canada}

\author{Guilu Long}
\affiliation{State Key Laboratory of Low-Dimensional Quantum Physics and Department of Physics, Tsinghua University, Beijing 100084, China}
\affiliation{Tsinghua National Laboratory of Information Science and Technology and The Innovative Center of Quantum Matter, Beijing 100084, China}
\affiliation{Beijing Academy of Quantum Information Sciences, Beijing 100193, China}

\author{Bei Zeng}
\email{zengb@uoguelph.ca}
\affiliation{Department of Mathematics \& Statistics, University of Guelph, Guelph N1G 2W1, Ontario, Canada}%
\affiliation{Institute for Quantum Computing, University of Waterloo, Waterloo N2L 3G1, Ontario, Canada}
\affiliation{Shenzhen Institute for Quantum Science and Engineering and Department of Physics, Southern University of Science and Technology, Shenzhen 518055, China}

\begin{abstract}
Quantum state tomography is a daunting challenge of experimental quantum computing even in moderate system size. One way to boost the efficiency of state tomography is via local measurements on reduced density matrices, but the reconstruction of the full state thereafter is hard. Here, we present a machine learning method to recover the full quantum state from its local information, where a fully-connected neural network is built to fulfill the task with up to seven qubits. In particular, we test the neural network model with a practical dataset, that in a 4-qubit nuclear magnetic resonance system our method yields global states via the 2-local information with high accuracy. Our work paves the way towards scalable state tomography in large quantum systems.
\end{abstract}

\maketitle

\section{Introduction}
Quantum state tomography (QST) plays a vital role in validating and benchmarking quantum devices \cite{d2002quantum, haffner2005scalable, leibfried2005creation, lvovsky2009continuous, baur2012benchmarking} because it can completely capture properties of an arbitrary quantum state. However, QST is not feasible for large systems because of its need for exponential resources. In recent years, there has been extensive research on methods for boosting the efficiency of QST  \cite{klimov2008optimal,hou2016full,cramer2010efficient,gross2010quantum,toth2010permutationally,li2017optimal,lanyon2017efficient}. One of the promising candidates among these methods is QST via reduced density matrices (RDMs) \cite{kalev2015power, linden2002almost, linden2002parts, diosi2004three, chen2012comment, chen2012ground, chen2013uniqueness}; because local measurements are convenient and accurate on many experimental platforms.


QST via RDMs is also a useful tool for characterizing ground states of local Hamiltonians. A many-body Hamiltonian $H$ is $k$-local if $H=\sum_i H_i^{(k)}$, where each term $H_i^{(k)}$ acts non-trivially on at most $k$ particles. In practical situations, we would mainly be interested in $2$-local Hamiltonians, sometimes with certain interaction patterns, such as nearest-neighbor interactions on some lattices. In this case, $k$-RDMs of a $k$-local Hamiltonian uniquely determine its unique ground state~\cite{chen2012ground}. Therefore, for these ground states, one only needs $k$-local measurements (i.e., $k$-RDMs) for state tomography.

Although local measurements are efficient, in general, reconstructing the state from the measurement results is known to be hard~\cite{qi2013quantum}. To be more precise, it is easy to obtain the $k$-RDMs of any $n$-particle quantum state $\ket{\psi}$. However, even if $\ket{\psi}$ is uniquely determined by its $k$-RDMs, reconstructing $\ket{\psi}$ from its $k$-RDMs is computationally hard. We remark that this is not due to the problem that $\ket{\psi}$ needs to be described by exponentially many parameters. In fact, in many cases, ground states of $k$-local Hamiltonians can be effectively represented by tensor product states~\cite{zeng2015quantum}.

\begin{figure}[h]
    \centering
    \includegraphics[width=0.8\linewidth]{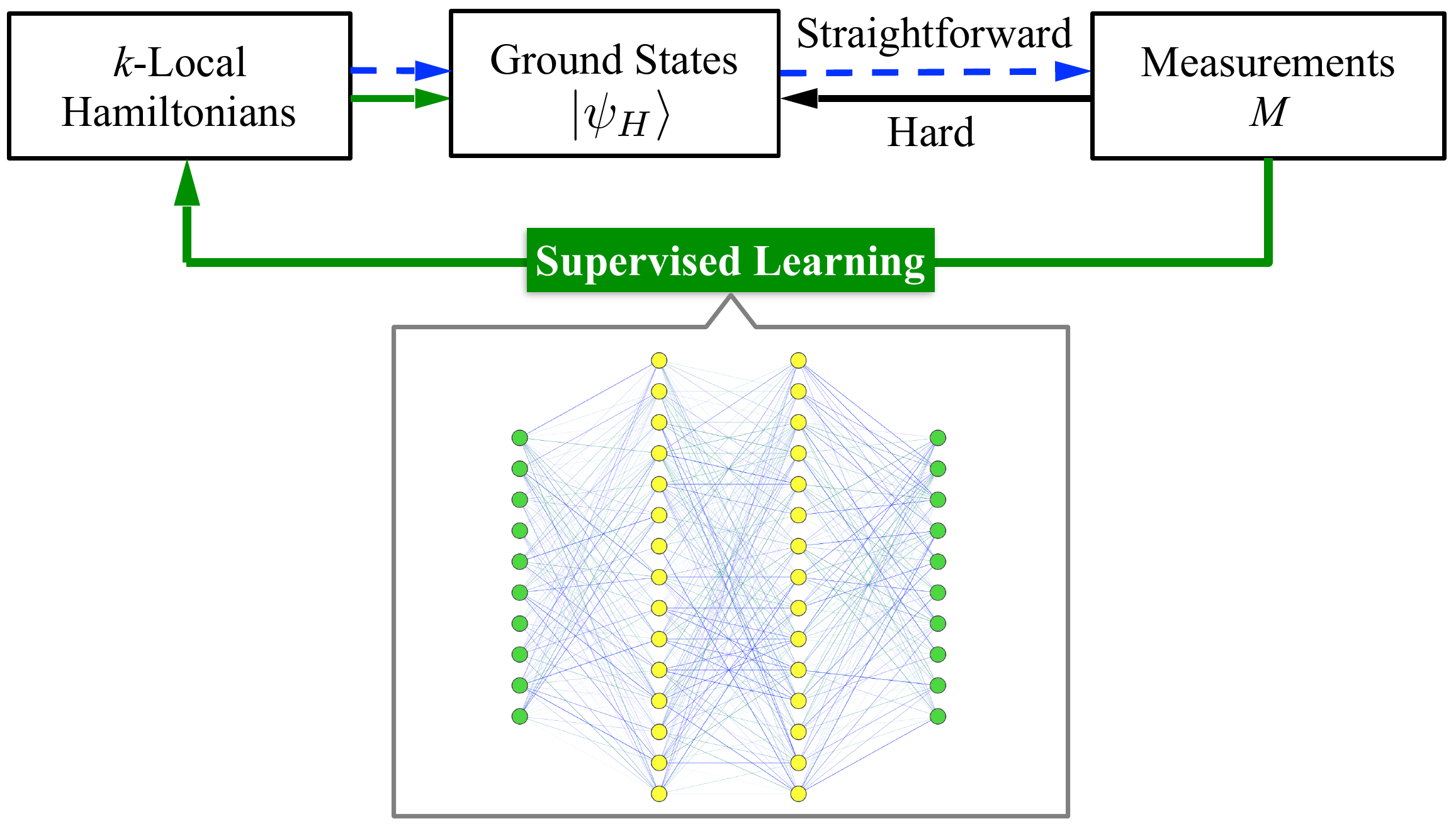}
    \caption{\textbf{Procedure of our neural network based local quantum state tomography method.} As shown by the blue dashed arrows, we first construct training and test dataset by generating random $k$-local Hamiltonians $H$, calculate their ground states $|\psi_H|\rangle$, and obtained local measurement results $M$. We then train the neural network with the generated training dataset. After training, as represented by the green arrows, we first obtain the Hamiltonian $H$ through local measurement results $\textbf{M}$ from the neural network, then recover the ground states from the obtained Hamiltonian. In contrast, the black arrow represents the direction of the normal QST process, which is computationally hard.}
    \label{process}
\end{figure}

The state reconstruction problem naturally connects to the regression problem in supervised learning. Regression analysis, in general, seeks to discover the relation between inputs and outputs, i.e., to recover the underlying mathematical model. Unsupervised learning techniques have been applied to QST in various cases, such as in Refs. \cite{kieferova2017tomography, torlai2018neural}. In our case, as shown in \cref{process}, by knowing the Hamiltonian $H$, it is relatively easy to get the ground state $\ket{\psi_H}$ since the ground state is nothing but the eigenvector corresponds to the smallest eigenvalue. And then we could naturally achieve the $k$-local measurements $\mathbf{M}$ of $\ket{\psi_H}$. Therefore, the data for tuning our reverse engineering model is accessible, which allows us to realize QST through supervised learning practically.


In this work, we proposed a local-measurement-based QST by fully-connected feedforward neural network, in which every neuron connects to every neuron in the next layer and information only passes forward (i.e., have no loop in the network). We first build a fully-connected feedforward neural network for $4$-qubit ground states of fully-connected $2$-local Hamiltonians. Our trained $4$-qubit network not only predicts the test dataset with high fidelity but also reconstruct $4$-qubit nuclear magnetic resonance (NMR) experimental states accurately. We use the $4$-qubit case to demonstrate the potential of using neural networks to realize QST via RDMs. The versatile framework of neural networks for recovering ground states of $k$-local Hamiltonians could be extended to more qubits and various interaction structures; we then apply our methods to the ground states of seven-qubit 2-local Hamiltonians with nearest neighbors couplings and the ground states of translational invariant 2-local Hamiltonians up to 15 qubits. In both cases, neural networks give accurate predictions with high fidelities.


\section{Results}

\subsection{Theory}

\begin{figure}[t]
  \centering 
  \subfloat[\label{fig:conf1}]{\includegraphics[width=0.3\linewidth]{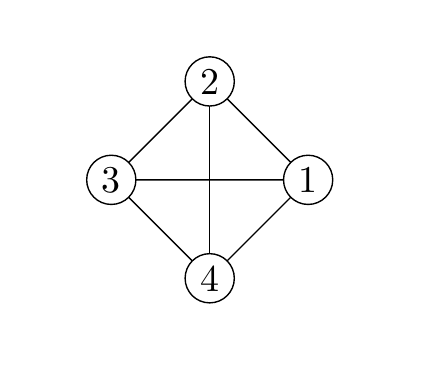}}
  \hspace{0.5in} 
  \subfloat[\label{fig:conf2}]{\includegraphics[width=0.5\linewidth]{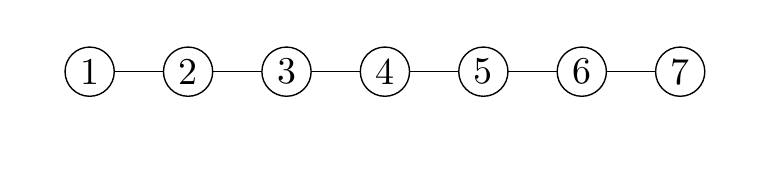}}\\
  \subfloat[\label{fig:4f}]{\includegraphics[width=0.5\linewidth]{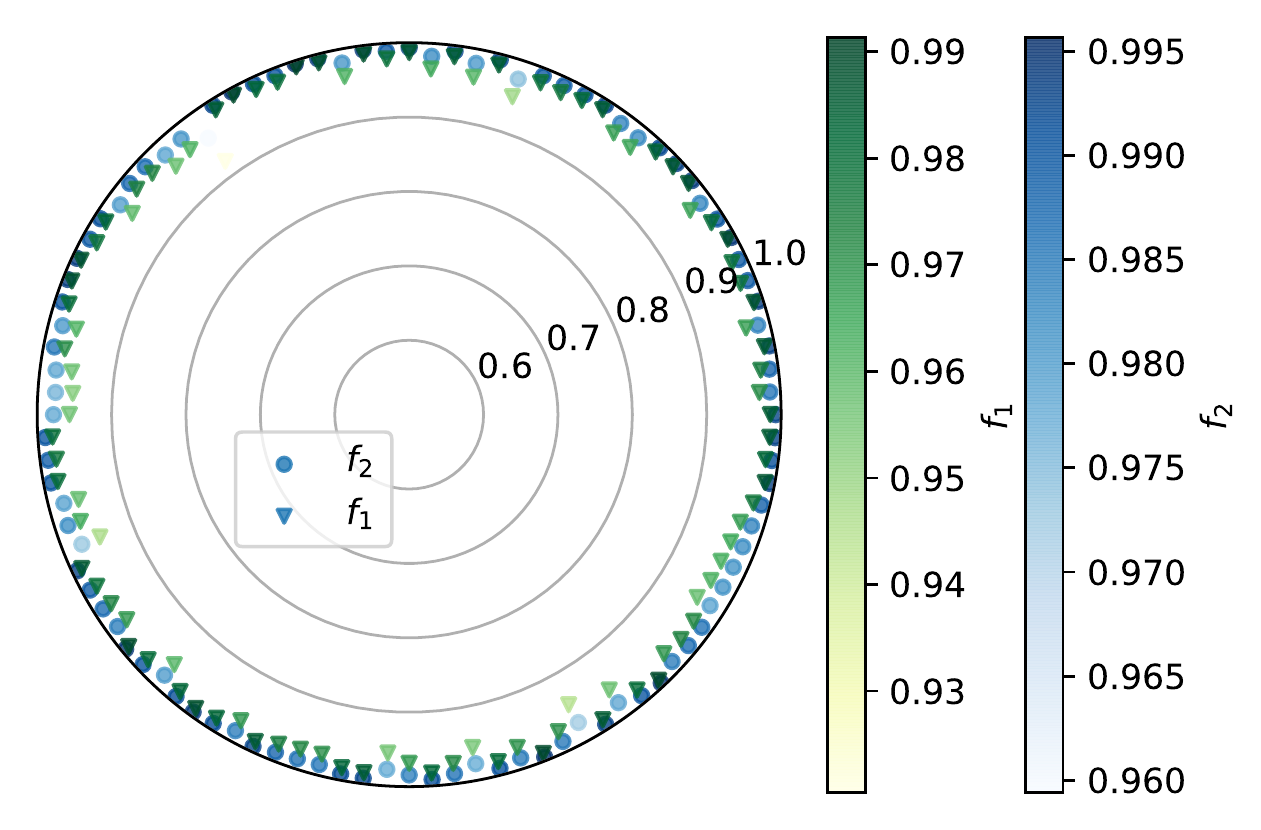}}
  \subfloat[\label{fig:7f}]{\includegraphics[width=0.5\linewidth]{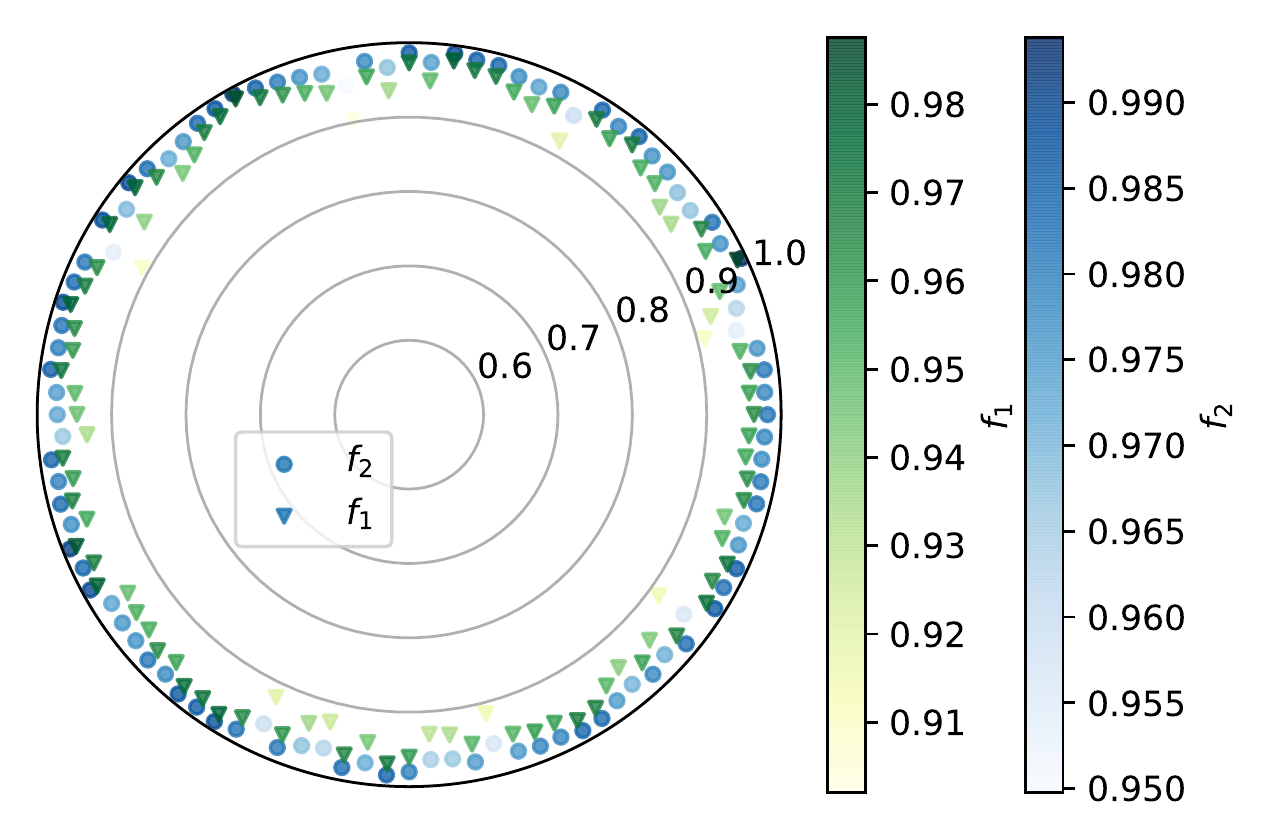}}
  \caption{\textbf{Theoretical results for 4 qubits and 7 qubits.} (a) The configuration of our 4-qubit states. Each dot presents a qubit, and every qubit interacts with each other. (b) The $f_1$ and $f_2$ of 100 random 4-qubit states $\rho_{\text{rd}}$ and our neural network predictions $\rho_{\text{nn}}$. (c) The configuration of our 7-qubit states: only nearest qubits have interactions. (d) The $f_1$ and $f_2$ of 100 random 7-qubit states $\rho_{\text{rd}}$ and our neural network predictions $\rho_{\text{nn}}$. } 
  \label{f} 
\end{figure}

The universal approximation theorem \citep{le2008representational} states that every continuous function on the compact subsets of $\mathbb{R}^n$ can be approximated by a multi-layer feedforward neural network with a finite number of neurons, i.e., computational units. And by observing the relation between $k$-local Hamiltonian and local measurements of its ground state, as shown in \cref{process}, we are empowered to turn the tomography problem to a regression problem which fit perfectly into the neural network framework.

In particular, we first construct a deep neural network for $4$-qubit ground states of full $2$-local Hamiltonians as follows:
\begin{equation}
H =  \sum \limits_{i = 1}^4 \sum \limits_{1\le k \le 3} \omega_k^{(i)} \sigma_k^{(i)}+ \sum \limits_{1\le i<j \le 4}\sum \limits_{1\le n, m \le 3} J_{nm}^{(ij)}\sigma_n^{(i)}\otimes \sigma_m^{(j)},
\label{ham}
\end{equation}
where $\sigma_{k}, \sigma_n, \sigma_m \in \Delta$, and $ \Delta= \{\sigma_1=\sigma_x, \sigma_2=\sigma_y, \sigma_3=\sigma_z, \sigma_4 = I\}$. 
We denote the set of Hamiltonian coefficients as $\vec h = \{\omega_k^{(i)}, J_{nm}^{(ij)}\}$. The coefficient vector $\vec h$ is the vector representation of $H$ according to the basis set $\mathbf{B} = \{\sigma_m \otimes \sigma_n: n+m \ne 8, \sigma_m,\sigma_n \in \Delta \}$. The configuration of the ground states is illustrated in \cref{fig:conf1}.

The number of parameters of the local observables $\mathbf{M}$ of ground states determines the amount of network input units. Concretely, $\mathbf{M} = \{s_{m,n}^{(i,j)}: s_{m,n}^{(i,j)} = \Tr(\Tr_{(i,j)}\rho \cdot B_{(m,n)}), B_{(m,n)}\in \mathbf{B},1 \le i < j \le 4, 1\le n, m \le 4 \}$, where $\sigma_n, \sigma_m \in \Delta$ and $\rho$ is the density matrix of the ground state. The input layer has $66$ neurons since the cardinality of the set of measurement results is $66$. Our network then contains two fully connected hidden layers, in which every neuron in the previous layer is connected to every neuron in the next layer. The number of output units equals to the number of parameters of our $2$-local Hamiltonian, which is $66$ in our $4$-qubit case. More details of our neural network can be found in Methods section.

Our training data consist of the 120,000 randomly generated $2$-local Hamiltonians as output and the local measurements of their corresponding ground states. The test data include 5,000 pairs of Hamiltonians and local measurement results $(H_i, \mathbf{M}_i)$.

We train the network by a popular optimizer in the machine learning community called Adam (Adaptive Moment Estimation) \cite{kingma2014adam,reddi2018convergence}. For loss function, we choose cosine proximity $ \cos (\theta) = (\vec h_{\text{pred}} \cdot \vec h)/(\|\vec h_{\text{pred}}\| \cdot \|\vec h\|)$, where $\vec h_{\text{pred}}$ is the prediction of the neural network and $\vec h$ is the desired output. We find the cosine proximity function fits our scenario better than the more commonly chosen loss functions such as mean square error or mean absolute error. The reason can be understood as follow. Note the parameter vector $\vec h$ is the representation of corresponding Hamiltonian in the Hilbert space expanded by the local operators $\mathbf{B}$. The angle $\theta$ between two vectors is a distance measure between two corresponding Hamiltonians \cite{qi2017determining}.

As illustrated in \cref{process}, after getting predicted Hamiltonian from the neural network, we calculate the ground state $\rho_{\text{nn}}$ of the predicted Hamiltonian and take the result as the prediction of ground state that we attempt to recover. We remark that our predicted Hamiltonian is not necessarily exactly the same as the original Hamiltonian; Even if that happens, our numeric results suggest their ground states are still close.

We use two different fidelities to measure the distance between the randomly generated states $\rho_{\text{rd}}$ and our neural network predicted states $\rho_{\text{nn}}$:
\begin{align}
\label{f1}
f_1(\rho_1, \rho_2) &\equiv \frac{\text{Tr}(\rho_1 \rho_2)}{\sqrt{\text{Tr}(\rho_1^2})\cdot\sqrt{\text{Tr}(\rho_2^2})},\\
\label{f2}
f_2(\rho_1, \rho_2)  &\equiv \text{Tr}\sqrt{\sqrt{\rho_1}\rho_2\sqrt{\rho_1}}.
\end{align}
Although the fidelity measure $f_2$ defined in \cref{f2} is standard \cite{nielsen2002quantum}, in experiments the measure $f_1$ are more convenient because it does not require the density matrix $\rho$ to be positive definite; in NMR experiments, the density matrix obtained directly from the raw data of a state tomography experiment may not be positive definite. Thus, we use $f_1$ for between any two pair of $\rho_{\text{nn}}$, theoretical state $\rho_{\text{th}}$ and experimental states $\rho_{\text{ml}}$ in the experiment section.

\begin{table}[h] 
\begin{center}
\begin{tabular}[c]{|c||c|c||c|c||c|c||c|c|}
\hline
& \multicolumn{2}{c||}{Max} & \multicolumn{2}{c||}{Min} & \multicolumn{2}{c||}{Standard Deviation} & \multicolumn{2}{c|}{Average Fidelity} \\\cline{2-9}
& $f_1$&$f_2$&$f_1$&$f_2$&$f_1$&$f_2$&$f_1$&$f_2$ \\ \hline
4-qubit & 99.6\% & 99.8\% & 83.5\% & 91.4\% & 11.6e-3 & 5.93e-3 & 97.6\% & 98.8\% \\ \hline
7-qubit & 99.2\% & 99.6\% & 72.7\% & 85.2\% & 20.2e-3 & 10.4e-3 & 95.9\% & 97.9\% \\\hline
\end{tabular}
\caption{\textbf{The statistical performance of our neural networks for 4-qubit and 7-qubit cases.}}
\label{tab:stats}
\end{center}
\end{table}

After supervised learning on the training data, our neural network is capable of predicting the 4-qubit output of the test set with high performance. The fidelity average over the whole test set is 97.5\% for $f_1$ and 98.7\% for $f_2$. The maximum, minimum, standard deviation of fidelities for the test set show in \cref{tab:stats}. \cref{fig:4f} illustrates the two fidelities between 100 random states $\rho_{\text{rd}}$ and our neural network predictions $\rho_{\text{nn}}$.

Our framework generalizes directly to more qubits and different interaction patterns. We apply our framework to recover 7-qubit ground states of $2$-local Hamiltonians with nearest neighbor interaction. The configuration of our $7$-qubit states shows in \cref{fig:conf2}. The Hamiltonian of this 7-qubit case is 

\begin{equation}
H =  \sum \limits_{i = 1}^7 \sum \limits_{1\le k \le 3} \omega_k^{(i)} \sigma_k^{(i)}+ \sum \limits_{i=1}^6\sum \limits_{1\le n, m \le 3} J_{nm}^{(i)}\sigma_n^{(i)}\otimes \sigma_m^{(i+1)},
\label{ham7}
\end{equation}
where $\sigma_{k}, \sigma_n, \sigma_m \in \Delta$, $\omega_k^{(i)}$ and $J_{nm}^{(i)}$ are coefficients. We trained a similar neural network with 250,000 pairs of random generated Hamiltonians and $2$-local measurements of corresponding ground states. The network predicts the 5,000 randomly generated test set with fidelity $f_1$ of 95.9\% and fidelity $f_2$ of 97.9\%. More statistical performance shows in \cref{tab:stats} and fidelity results of 100 random generated states show in \cref{fig:7f}.



\subsection{Experiment}

\begin{figure}[h]
\centering
\includegraphics[width=0.8\textwidth]{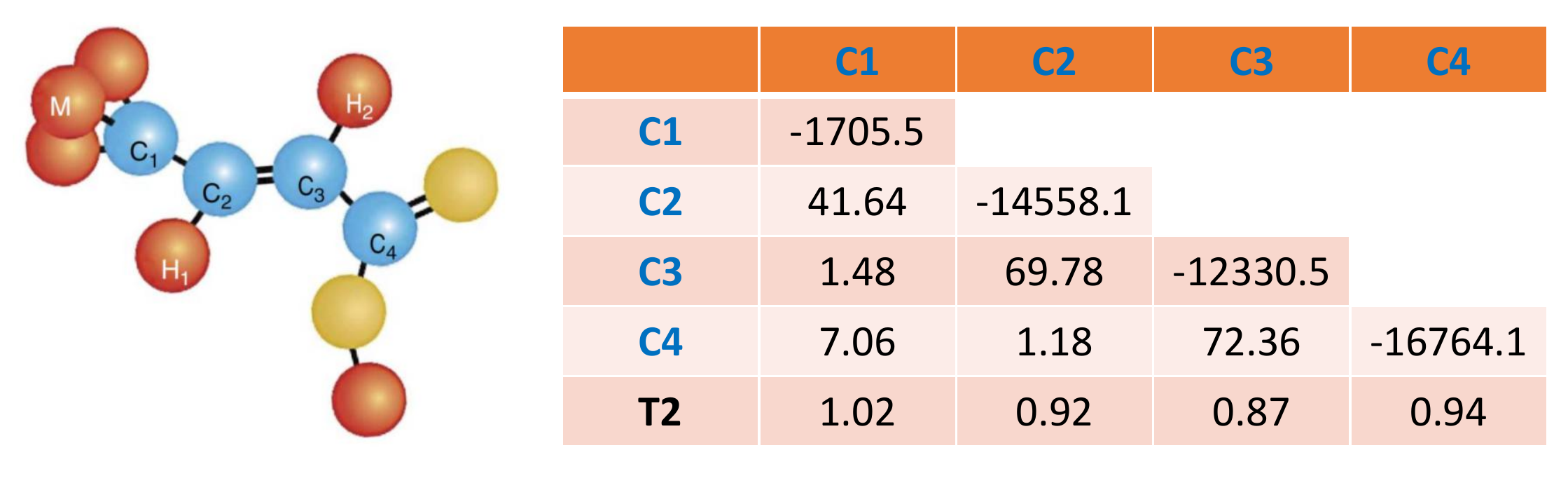}
\caption{\textbf{The molecular structure and Hamiltonian parameters of the $^{13}$C-labeled trans-crotonic acid}. The atoms C$_1$, C$_2$, C$_3$ and C$_4$ are used as the four qubits in the experiment, and the atoms M, H$_1$ and H$_2$ are decoupled throughout the experiment. In the table, the chemical shifts with respect to the Larmor frequency and J-coupling constants (in Hz) are listed by the diagonal and off-diagonal numbers, respectively. The relaxation timescales $T_{2}$ (in seconds) are shown at the bottom.} \label{molecule}
\end{figure}

So far, our theoretical model is noise-free. To demonstrate our trained machine learning model is resilient to experimental noises, we experimentally prepare the ground states of the random Hamiltonians and then try to reconstruct the final quantum states from 2-local RDMs using a four-qubit nuclear magnetic resonance (NMR) platform \cite{xin2018nuclear,vandersypen2005nmr,jones2000geometric,xin2018nmrcloudq}. The four-qubit sample is $^{13}$C-labeled trans-crotonic acid dissolved in d6-acetone, where C$_1$ to C$_4$ are encoded as the four work qubits, and the rest spin-half nuclei are decoupled throughout all experiments. \cref{molecule} describes the parameters and structure of this molecule. Under the weak-coupling approximation, the Hamiltonian of the system writes,
\begin{equation}
\mathcal{H}_{\text{int}}=\sum\limits_{j=1}^4 {\pi (\nu _j-\nu _0) } \sigma_z^j  + \sum\limits_{j < k,=1}^4 {\frac{\pi}{2}} J_{jk} \sigma_z^j \sigma_z^k,
\label{Hamiltonian}
\end{equation}
where $\nu_j$ are the chemical shifts, $\emph{J}_{jk}$ are the J-coupling strengths, and $\nu _0$ is the reference frequency of $^{13}$C channel in the NMR platform. All experiments were carried out on a Bruker AVANCE 400 MHz spectrometer at room temperature. We briefly describe our three experiments steps here and leave the details in the Methods section: (i) Initialization. The pseudo-pure state \cite{cory1997ensemble,fahmy2000nuclear,knill1998effective} for being the input of quantum computation $\ket{0000}$ is prepared. (More details are provided in the Methods section) (ii) Evolution. Starting from the state $\ket{0000}$, we create the ground state of the random two-body Hamiltonian by applying the optimized shaped pulses. (iii) Measurement. We measure the two-body reduced density matrices and perform four-qubit quantum state tomography (QST) to estimate the quality of our implementations.

\begin{figure}[h]
    \includegraphics[width=0.8\linewidth]{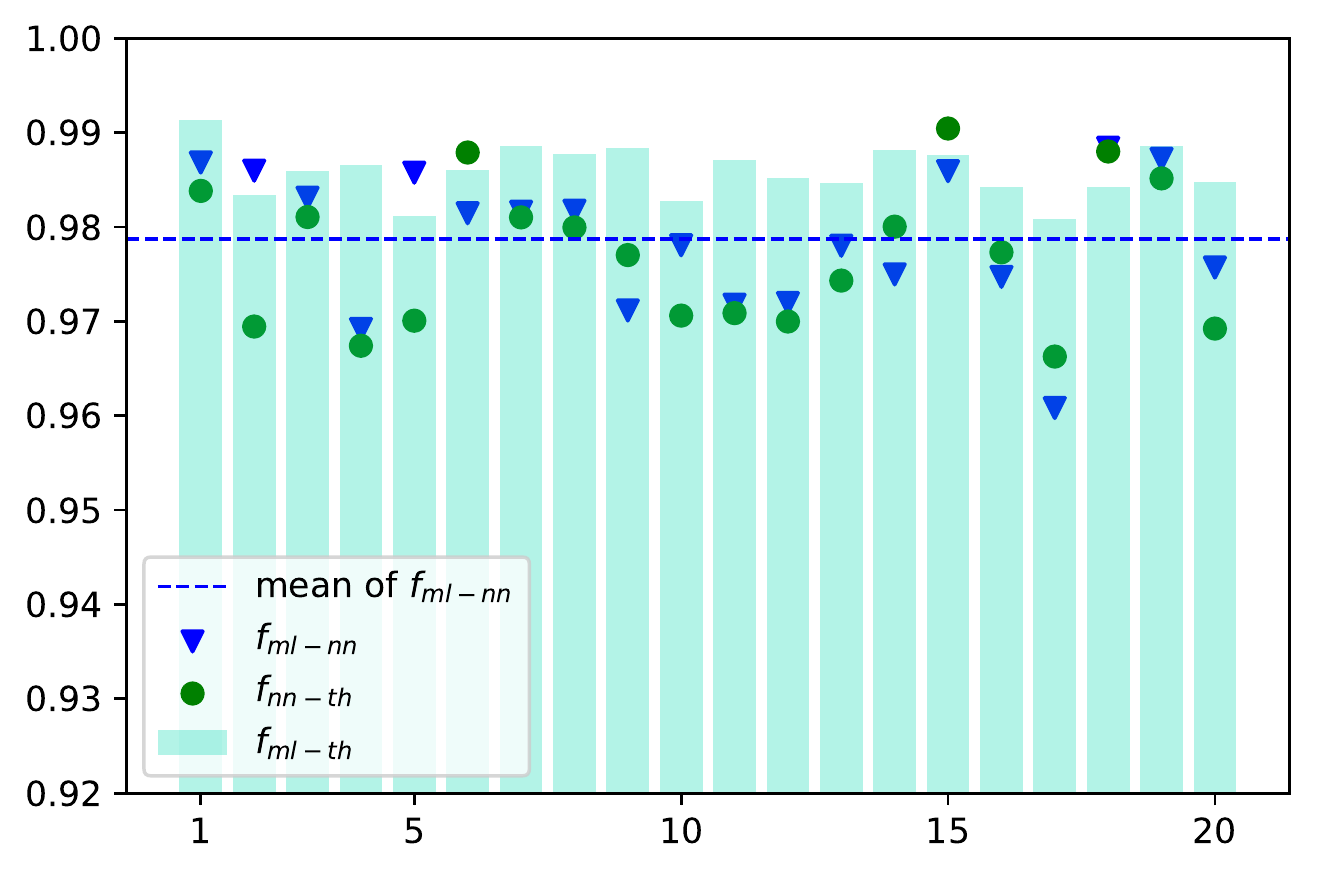}
    \caption{\textbf{The predication results with experimental data.}  Here we list three different fidelities ($f_1$ \cref{f1}) for 20 experimental instances. The horizontal axis is the dummy label of the 20 experimental states. The cyan bars, $f_{\text{ml}-\text{th}}$, are the fidelities between the theoretical states $\rho_{\text{th}}$ and the experimental states $\rho_{\text{ml}}$. The blue triangles, $f_{\text{ml}-\text{nn}}$,  are fidelities between our neural network predictions $\rho_{\text{nn}}$ and the experimental states $\rho_{\text{ml}}$ with the average fidelity over 97.9\%. And the green dots, $f_{\text{nn}-\text{th}}$, are the fidelities between our neural network predictions and the theoretical states. } 
    \label{fidelity}
\end{figure}

In experiments, we created the ground states of 20 random Hamiltonians of the form in \cref{ham} and performed $4$-qubit QST for them after the state preparations. First, we report that the average fidelities between the experimental states $\rho_{\text{ml}}$ (Note the subscript ml denotes a standard tomography method called maximum likelihood, rather than machine learning) and the target ground state $\rho_{\text{th}}$ is about 98.2\%. Second, we used $2$-RDMs of these density matrices to reconstruct $4$-qubit states by our neural-network-based framework, obtaining a average fidelity $f_1(\rho_{\text{ml}}, \rho_{\text{nn}})$ of 97.9\%, where $\rho_{\text{nn}}$ is the neural network predicted state. \cref{fidelity} shows the fidelity details of these density matrices. The results indicate that the original $4$-qubit state can be efficiently reconstructed by our trained neural network using only $2$-RDMs, instead of the traditional full QST.

\section{Methods}

\subsection{Machine Learning}

In this subsection, we discuss our training/test dataset generation procedure, the structure, and hyperparameters of our neural network, and the required number of training data during training.

The training and test data sets are formed by random $k$-local Hamiltonians and $k$-local measurements of corresponding ground states. For our $4$-qubit case, $2$-local Hamiltonians as defined in \cref{ham}. The parameter vector $\vec h$ of random Hamiltonians are uniformly drawn from random normal distributions without uniform mean values and standard deviations. It realized by applying function \verb|np.random.normal| in Python. Similarly, for the 7-qubit case, Hamiltonian is defined in \cref{ham7}, and the corresponding parameter vector $\vec h$ is generated by the same method. As the blue dashed lines in \cref{process} shown, after getting random Hamiltonians $H$, we calculate the ground states $\ket{\psi_H}$ (the eigenket corresponds to the smallest eigenvalue of $H$) and then get the $2$-local measurements $\mathbf{M}$. 


In this work, we use a fully-connected feedforward neural network, which is famous as the first and most simple type of neural network \cite{schmidhuber2015deep}. By fully-connected, it means every neuron is connected to every other neuron in the next layer. Feedforward or acyclic, as the word indicated, means information only passes forward; the network has no cycle. Our machine learning process is implemented using Keras \cite{chollet2015keras} which is a high-level deep learning library running on top of the popular machine learning framework: Tensorflow \cite{abadi2016tensorflow}.

As mentioned in the results section, experimental accessible data have been used as input to our neural network. The input is $\mathbf{M} = \{s_{m,n}^{(i,j)}: s_{m,n}^{(i,j)} = \Tr(\Tr_{(i,j)}\rho \cdot B_{(m,n)}), B_{(m,n)}\in \mathbf{B},1 \le i < j \le 4, 1\le n, m \le 4 \}$. For the $4$-qubit case, it is easy to see that $\mathbf{M}$ has $3\times 4 = 12$ single body terms and $C_4^2\times 9 = 54$ $2$-body terms. By arranging these $66$ elements in $\mathbf{M}$ into a row, we set it as the input of our neural network. 
The output set to be the vector representation of the Hamiltonian $\vec h$, which also has 66 entries. 
For the $7$-qubit $2$-local case, where $2$-body terms only appear on nearest qubits, the network takes $2$-local measurements as input, and the number of neurons in the input layer is $7\times 3 + 6\times 3\times 3 = 75$. The number of neurons in the output layer is also 75.

The physical aspect of our problem fixes the input and output layers. The principle for setting hidden layers leads by efficiency. While training networks, inspired by Occam's Razor principle, we choose fewer layers and neurons when increasing them do not significantly increase the performance but increases the required training epochs. In our $4$-qubit case, two hidden layers of 300 neurons have been inserted between the input layer and the output layer. In the $7$-qubit case, we use four fully-connected hidden layers with the following number of hidden neurons: 150-300-300-150. The activate function for each layer is ReLU (Rectified Linear Unit) \cite{nair2010rectified}, which is a widely used non-linear activation function. We also choose the optimizer having the best performance in our problem over almost all the built-in optimizers in Tensorflow: AdamOptimizer (Adaptive Moment Estimation) \cite{kingma2014adam}. The learning rate is set to be 0.001.

\begin{table}[h]
\begin{center}
\begin{tabular}[c]{|c||c|c||c|c||c|c|}
\hline
\multicolumn{7}{|c|}{\textbf{4-qubit (66-300-300-66)}}\\ \hline
 & \multicolumn{2}{c||}{epoch:100}& \multicolumn{2}{c||}{epoch:300}& \multicolumn{2}{c|}{epoch:600} \\ \cline{2-7}
 Training data& $f_1$&$f_2$&$f_1$&$f_2$&$f_1$&$f_2$ \\ \hline
500 & 50.5\% & 69.1\% &  35.5\% & 56.2\% & 34.2\% & 56.9\% \\ \hline
1,000 & 62.6\% & 78.1\% &  47.6\% & 66.8\% &  38.3\% & 58.7\% \\ \hline
5,000 & 89.7\% & 94.7\% &  88.5\% & 94.1\% & 87.8\% & 93.6\% \\ \hline
10,000 & 93.3\% & 96.6\% & 93.2\% & 96.5\% & 93.1\% & 96.5\% \\ \hline
50,000 & 96.1\% & 98.0\% & 96.8\% & 98.4\% & 96.3\% & 98.1\% \\ \hline
100,000 & 96.8\% & 98.4\% & 97.2\% & 98.6\% & 97.0\% & 98.5\% \\\hline
120,000 & 97.0\% & 98.5\% & \cellcolor[rgb]{.7,.9,.9} 97.3\% & \cellcolor[rgb]{.7,.9,.9} 98.6\% & 97.1\% & 98.6\% \\ \hline

\multicolumn{7}{|c|}{\textbf{7-qubit (75-150-300-300-150-75)}}\\ \hline
1,000 & 13.6\% & 22.2\% & 12.5\% & 31.2\% &  12.1\% & 31.3\% \\ \hline
10,000 & 88.0\% & 93.8\% & 83.9\% & 91.5\% &  78.9\% & 88.7\% \\ \hline
50,000 & 93.1\% & 96.5\% & 93.4\% & 96.6\% &  93.8\% & 96.9\% \\ \hline
100,000 & 93.9\% & 96.9\% & 94.7\% & 97.3\% & 95.2\% & 97.5\% \\\hline
150,000 & 94.3\% & 97.1\% & 95.5\% & 97.7\% & 95.5\% & 97.7\% \\\hline
200,000 & 94.7\% & 97.3\% & 95.6\% & 97.7\% & 95.7\% & 97.8\%  \\\hline
250,000 & 95.4\% & 97.6\% &\cellcolor[rgb]{.7,.9,.9}  95.7\% &\cellcolor[rgb]{.7,.9,.9} 97.8\% & 95.7\% & 97.8\% \\\hline

\end{tabular}
\caption{\textbf{Average fidelities on the test set by using different numbers of training data and epochs. } The batch size is 512, and the size of test dataset is 5000. As the amount of training data increases, we find the average fidelity of predicted states and the true test states goes up, and the neural network reaches a certain performance after we fed sufficient training data. We also observe more training data requires more training epochs;}
\label{tab:fid}
\end{center}
\end{table}

\begin{table}[h]
\begin{center}
\begin{tabular}[c]{|c||c|c||c|c||c|c|}
\hline
\multicolumn{7}{|c|}{\textbf{4-qubit (Training Data: 120,000)}}\\ \hline
 & \multicolumn{2}{c||}{epoch:300}& \multicolumn{2}{c||}{epoch:600}& \multicolumn{2}{c|}{epoch:900} \\ \cline{2-7}
 Batch Size& $f_1$&$f_2$&$f_1$&$f_2$&$f_1$&$f_2$ \\ \hline
512 &  97.3\% & 98.6\% & 97.1\% & 98.6\% & 97.5\% & 98.7\% \\ \hline
1028 & 97.3\% & 98.7\% &\cellcolor[rgb]{.7,.9,.9} 97.5\% & \cellcolor[rgb]{.7,.9,.9} 98.7\% & 97.5\% & 98.7\% \\ \hline
2048 & 97.0\% & 98.5\% & 97.4\% & 98.7\% & 97.5\% & 98.7\% \\\hline

\multicolumn{7}{|c|}{\textbf{7-qubit (Training Data: 250,000)}}\\ \hline
512 & 95.7\% &97.8\% & 95.7\% & 97.8\% &\cellcolor[rgb]{.7,.9,.9} 95.9\% & \cellcolor[rgb]{.7,.9,.9}97.9\% \\\hline
1028 & 95.3\% & 97.6\% & 95.6\% & 97.8\% & 95.7\% & 97.8\% \\ \hline
\end{tabular}
\caption{\textbf{Average fidelities on the test set by using different batch sizes.} The size of test dataset is 5,000. The optimal batch size for the 4-qubit case is 1024 and for the 7-qubit case is 512.}
\label{tab:delicate}
\end{center}
\end{table}

The whole training dataset has been split into two parts, 80\% used for training, and 20\% used for validation after each epoch. A new data set of 5,000 data was used as the test set after training. The initial batch size was chosen as 512. As the amount of training data increases, the average fidelity of predicted states and the true test states goes up. The neural network reaches a certain performance after we fed sufficient training data. More training data requires more training epochs; however, replete epochs ebb the neural network performance due to over-fitting. \cref{tab:fid} shows the average fidelities of using different training data and epochs. The first round of training locks down the optimal amount of training data, then we change the batch size and find the optimal epoch. We report the results for the second round training in \cref{tab:delicate}. For the 4-qubit case, appropriately increases the batch size can benefit the stability of training process thus improves the performance of the neural network. Though, by choosing the batch size as 512 and 2048, the network can also reach the same performance with larger epochs, we chose the batch size as 1028 since more epochs require more training time. After the same attempting for the 7-qubit case, we find 512 is promising batch size.


\subsection{NMR states preparation}

Our experiment procedure consists of three steps:  initialization, evolution, and measurement. In this subsection, we discuss these three steps in details.

(i) \textbf{Initialization.} The computational basis state $\ket{0}^{\otimes n}$ is usually chosen as the input state for quantum computation. Most of the quantum systems do not start from such an input state, so the special initialization processing is necessary before applying quantum circuits. In NMR, the sample initially stays in the Boltzmann distribution at room temperature,
\begin{align}\label{thermalstate}
\rho_{\text{thermal}}=\mathcal{I}/16+\epsilon(\sigma^1_z+\sigma^2_z+\sigma^3_z+\sigma^4_z),
\end{align}
where $\mathcal{I}$ is the $16\times 16$ identity matrix and $\epsilon=10^{-5}$ is the polarization. We can not directly use it as the input state for quantum computation, because such a thermal state is a highly-mixed state \cite{gershenfeld1997bulk,cory1997ensemble}. We instead create a so-called pseudo-pure state (PPS) from this thermal state by using the spatial averaging technique \cite{cory1997ensemble,fahmy2000nuclear,knill1998effective}, which consists of applying local unitary rotations and using $z$-gradient fields to destroy the unwanted coherence. The form of teh 4-qubit PPS can be wrote as
\begin{align}\label{PPS}
\rho_{0000}=(1-\epsilon')\mathcal{I}/16+\epsilon'\ket{0000}\bra{0000}.
\end{align}
Here, although the PPS $\rho_{0000}$ is also a highly-mixed state, the identity part $\mathcal{I}$ neither evolves under any unitary operations nor influences the experimental signal. It means that we can focus on the deviated part $\ket{0000}\bra{0000}$ and consider $\ket{0000}\bra{0000}$ as the initial state of our quantum system. Finally, 4-qubit QST was performed to evaluate the quality of our PPS. We found that the fidelity between the perfect pure state $\ket{0000}$ and the experimentally measured PPS is about 98.7\% by the definition $f_1$. 
This sets a solid ground for the subsequent experiments.

 (ii) \textbf{Evolution.} In this step, we prepared the ground states of the given Hamiltonians using optimized pulses. The form of the considered Hamiltonian is chosen as \cref{ham}.
Here, the parameters $\omega_k^{(i)}$ and $J_{nm}^{(ij)}$ mean the chemical shift and the J-coupling strength, respectively. In experiments, we create the ground states of different Hamiltonians by randomly changing the parameter set $(\omega_k^{(i)}, J_{nm}^{(ij)})$. For the given Hamiltonian, the gradient ascent pulse engineering (GRAPE) algorithm \cite{boulant2003experimental,khaneja2005optimal,ryan2008liquid,lu2017enhancing} is adopted to optimize a radio-frequency (RF) pulse to realize the dynamical evolution from the initial state $\ket{0000}$ to the target ground state . The GRARE pulses are designed to be robust to the static field distributions and RF inhomogeneity, and the simulated fidelity is over $0.99$ for each dynamical evolution.

(iii) \textbf{Measurement.} In principle, we only need to measure the two-body reduced density matrices ($2$-RDMs) to determine the original $4$-qubit Hamiltonian through our trained network. Experimentally, we performed $4$-qubit QST, which naturally includes the 2-RDMs after preparing these states \cite{leskowitz2004state,lee2002quantum,PhysRevA.96.032307}, to evaluate the performance of our implementations.
Hence, we can estimate the quality of the experimental implementations by computing the fidelity between the target ground state $\rho_{\text{th}}=|\psi_{\text{th}}\rangle\langle\psi_{\text{th}}|$ and the experimentally reconstructed density matrix. Considering that the ground state of the target Hamiltonian should be real, the experimental density matrix should also be real. We use a maximum likelihood approach to reconstruct the most likely pure state $\rho_{\text{ml}}$ \cite{altepeter2005photonic}.
By reconstructing states $\rho_{\text{NN}}$ merely based on the experimental 2-RDMs, the performance of our trained neural network can be evaluated by comparing the states $\rho_{\text{ml}}$ with the states $\rho_{\text{nn}}$.

Finally, we attempt to evaluate the confidence of the expected results by analyzing the potential error sources in experiments. The infidelity of the experimental density matrix is mainly caused by some primary aspects in experiments, including decoherence effects, imperfections of the PPS preparation, and imprecision of the optimized pulses. From a theoretical perspective, we numerically simulate the influence of the optimized pulses and the decoherence effect of our qubits. Then we compare the fidelity computed in this manner with the ideal case to evaluate the quality of the final density matrix. As a numerical result, about 0.2\% infidelity was created on average and the 1.2\% error related to the infidelity of the initial state preparation. Additionally, other errors can also contribute to the infidelity such as imperfections in the readout pulses and spectral fitting.



\section{Discussion}


As a famous double-edged sword in experimental quantum computing, QST captures full information of quantum states on the one hand, while on the other hand, its implementation consumes a tremendous amount of resources. Unlike traditional QST that requires exponential many experiments with the growth of system size, the recent approach by measuring RDMs and reconstructing the full state thereafter opens up a new avenue to efficiently realize experimental QST. However, there is still an obstacle in this approach, that it is in general computationally hard to construct the full quantum state from its local information.

This is a typical problem empowered by machine learning. In this work, we apply the neural network model to solve this problem and demonstrate the feasibility of our method with up to seven qubits in the simulation. It should be noticed that 7-qubit QST in experiments is already a significant challenge in many platforms -- the largest QST to date is of 10 qubits in superconducting circuits, where the theoretical state is a GHZ state with rather simple mathematical form \cite{PhysRevLett.119.180511}. We further demonstrate that our method works well in a 4-qubit NMR experiment, thus validating its usefulness in practice. We anticipate this method to be a powerful tool in future QST tasks of many qubits due to its accuracy and convenience.

Our framework can be extended in several ways. First, we can consider excited states. As stated in the Results section, the Hamiltonian recovered by our neural network is not necessarily the original Hamiltonian, but their ground states are fairly close. We preliminarily examined eigenstates of predicted Hamiltonians. Although the ground states have considerable overlap, the excited states are not close to each other. It means, in this reverse engineering problem, ground states are numerically more stable than excited states. To recover excited states using our method, one may need to use more sophisticated neural networks such as convolutional neural network \cite{krizhevsky2012imagenet} (CNN) or Residual neural network \cite{he2016deep} (ResNet). Second, although we haven't include noise in the training and test data, our network predicts the experimental 4-qubit fully-connected 2-local states with high fidelities. This indicates our method has certain error tolerant ability. For future study, one can add different noise to the training and test data.


\noindent {\bf Data Availability.} All data and code needed to evaluate the conclusions are available from the corresponding authors upon reasonable request.

\noindent {\bf Acknowledgments.} We thank Yi Shen for helpful discussions.


\noindent {\bf Competing Interests.} The authors declare that they have no
competing financial interests.

\noindent {\bf Funding.} T.X. and G.L. are grateful to the following funding sources: the National Natural Science Foundation of China (11175094); National Basic Research Program of China (2015CB921002). J.L. is supported by the National Science Fund for Distinguished Young Scholars (11425523) and NSAF (U1530401). N.C. and B.Z. acknowledge the Natural Sciences and Engineering Research Council of Canada (NSERC), Canadian Institute for Advanced Research (CIFAR) and Chinese Ministry of Education (20173080024).
    

\begin{thebibliography}{51}
\providecommand{\natexlab}[1]{#1}
\providecommand{\url}[1]{\texttt{#1}}
\expandafter\ifx\csname urlstyle\endcsname\relax
  \providecommand{\doi}[1]{doi: #1}\else
  \providecommand{\doi}{doi: \begingroup \urlstyle{rm}\Url}\fi

\bibitem[D'Ariano et~al.(2002)D'Ariano, De~Laurentis, Paris, Porzio, and
  Solimeno]{d2002quantum}
G~Mauro D'Ariano, Martina De~Laurentis, Matteo~GA Paris, Alberto Porzio, and
  Salvatore Solimeno.
\newblock Quantum tomography as a tool for the characterization of optical
  devices.
\newblock \emph{Journal of Optics B: Quantum and Semiclassical Optics},
  4\penalty0 (3):\penalty0 S127, 2002.

\bibitem[H{\"a}ffner et~al.(2005)H{\"a}ffner, H{\"a}nsel, Roos, Benhelm,
  Chwalla, K{\"o}rber, Rapol, Riebe, Schmidt, Becher,
  et~al.]{haffner2005scalable}
Hartmut H{\"a}ffner, Wolfgang H{\"a}nsel, CF~Roos, Jan Benhelm, Michael
  Chwalla, Timo K{\"o}rber, UD~Rapol, Mark Riebe, PO~Schmidt, Christoph Becher,
  et~al.
\newblock Scalable multiparticle entanglement of trapped ions.
\newblock \emph{Nature}, 438\penalty0 (7068):\penalty0 643, 2005.

\bibitem[Leibfried et~al.(2005)Leibfried, Knill, Seidelin, Britton, Blakestad,
  Chiaverini, Hume, Itano, Jost, Langer, et~al.]{leibfried2005creation}
Dietrich Leibfried, Emanuel Knill, Signe Seidelin, Joe Britton, R~Brad
  Blakestad, John Chiaverini, David~B Hume, Wayne~M Itano, John~D Jost,
  Christopher Langer, et~al.
\newblock Creation of a six-atom ‘schr{\"o}dinger cat’state.
\newblock \emph{Nature}, 438\penalty0 (7068):\penalty0 639, 2005.

\bibitem[Lvovsky and Raymer(2009)]{lvovsky2009continuous}
Alexander~I Lvovsky and Michael~G Raymer.
\newblock Continuous-variable optical quantum-state tomography.
\newblock \emph{Reviews of Modern Physics}, 81\penalty0 (1):\penalty0 299,
  2009.

\bibitem[Baur et~al.(2012)Baur, Fedorov, Steffen, Filipp, Da~Silva, and
  Wallraff]{baur2012benchmarking}
M~Baur, A~Fedorov, L~Steffen, S~Filipp, MP~Da~Silva, and A~Wallraff.
\newblock Benchmarking a quantum teleportation protocol in superconducting
  circuits using tomography and an entanglement witness.
\newblock \emph{Physical review letters}, 108\penalty0 (4):\penalty0 040502,
  2012.

\bibitem[Klimov et~al.(2008)Klimov, Munoz, Fern{\'a}ndez, and
  Saavedra]{klimov2008optimal}
AB~Klimov, C~Munoz, A~Fern{\'a}ndez, and C~Saavedra.
\newblock Optimal quantum-state reconstruction for cold trapped ions.
\newblock \emph{Physical Review A}, 77\penalty0 (6):\penalty0 060303, 2008.

\bibitem[Hou et~al.(2016)Hou, Zhong, Tian, Dong, Qi, Li, Wang, Nori, Xiang, Li,
  et~al.]{hou2016full}
Zhibo Hou, Han-Sen Zhong, Ye~Tian, Daoyi Dong, Bo~Qi, Li~Li, Yuanlong Wang,
  Franco Nori, Guo-Yong Xiang, Chuan-Feng Li, et~al.
\newblock Full reconstruction of a 14-qubit state within four hours.
\newblock \emph{New Journal of Physics}, 18\penalty0 (8):\penalty0 083036,
  2016.

\bibitem[Cramer et~al.(2010)Cramer, Plenio, Flammia, Somma, Gross, Bartlett,
  Landon-Cardinal, Poulin, and Liu]{cramer2010efficient}
Marcus Cramer, Martin~B Plenio, Steven~T Flammia, Rolando Somma, David Gross,
  Stephen~D Bartlett, Olivier Landon-Cardinal, David Poulin, and Yi-Kai Liu.
\newblock Efficient quantum state tomography.
\newblock \emph{Nature communications}, 1:\penalty0 149, 2010.
\newblock URL \url{https://www.nature.com/articles/ncomms1147}.

\bibitem[Gross et~al.(2010)Gross, Liu, Flammia, Becker, and
  Eisert]{gross2010quantum}
David Gross, Yi-Kai Liu, Steven~T. Flammia, Stephen Becker, and Jens Eisert.
\newblock Quantum state tomography via compressed sensing.
\newblock \emph{Phys. Rev. Lett.}, 105:\penalty0 150401, Oct 2010.
\newblock \doi{10.1103/PhysRevLett.105.150401}.
\newblock URL \url{https://link.aps.org/doi/10.1103/PhysRevLett.105.150401}.

\bibitem[T{\'o}th et~al.(2010)T{\'o}th, Wieczorek, Gross, Krischek, Schwemmer,
  and Weinfurter]{toth2010permutationally}
G{\'e}za T{\'o}th, Witlef Wieczorek, David Gross, Roland Krischek, Christian
  Schwemmer, and Harald Weinfurter.
\newblock Permutationally invariant quantum tomography.
\newblock \emph{Physical review letters}, 105\penalty0 (25):\penalty0 250403,
  2010.

\bibitem[Li et~al.(2017{\natexlab{a}})Li, Huang, Luo, Li, Lu, and
  Zeng]{li2017optimal}
Jun Li, Shilin Huang, Zhihuang Luo, Keren Li, Dawei Lu, and Bei Zeng.
\newblock Optimal design of measurement settings for quantum-state-tomography
  experiments.
\newblock \emph{Physical Review A}, 96\penalty0 (3):\penalty0 032307,
  2017{\natexlab{a}}.

\bibitem[Lanyon et~al.(2017)Lanyon, Maier, Holz{\"a}pfel, Baumgratz, Hempel,
  Jurcevic, Dhand, Buyskikh, Daley, Cramer, et~al.]{lanyon2017efficient}
BP~Lanyon, C~Maier, M~Holz{\"a}pfel, T~Baumgratz, C~Hempel, P~Jurcevic,
  I~Dhand, AS~Buyskikh, AJ~Daley, M~Cramer, et~al.
\newblock Efficient tomography of a quantum many-body system.
\newblock \emph{Nature Physics}, 13\penalty0 (12):\penalty0 1158, 2017.
\newblock URL \url{https://www.nature.com/articles/nphys4244}.

\bibitem[Kalev et~al.(2015)Kalev, Baldwin, and Deutsch]{kalev2015power}
Amir Kalev, Charles~H Baldwin, and Ivan~H Deutsch.
\newblock The power of being positive: Robust state estimation made possible by
  quantum mechanics.
\newblock \emph{arXiv preprint arXiv:1511.01433}, 2015.

\bibitem[Linden et~al.(2002)Linden, Popescu, and Wootters]{linden2002almost}
N~Linden, S~Popescu, and WK~Wootters.
\newblock Almost every pure state of three qubits is completely determined by
  its two-particle reduced density matrices.
\newblock \emph{Physical review letters}, 89\penalty0 (20):\penalty0 207901,
  2002.

\bibitem[Linden and Wootters(2002)]{linden2002parts}
N~Linden and WK~Wootters.
\newblock The parts determine the whole in a generic pure quantum state.
\newblock \emph{Physical review letters}, 89\penalty0 (27):\penalty0 277906,
  2002.

\bibitem[Di{\'o}si(2004)]{diosi2004three}
Lajos Di{\'o}si.
\newblock Three-party pure quantum states are determined by two two-party
  reduced states.
\newblock \emph{Physical Review A}, 70\penalty0 (1):\penalty0 010302, 2004.

\bibitem[Chen et~al.(2012{\natexlab{a}})Chen, Ji, Ruskai, Zeng, and
  Zhou]{chen2012comment}
Jianxin Chen, Zhengfeng Ji, Mary~Beth Ruskai, Bei Zeng, and Duan-Lu Zhou.
\newblock Comment on some results of erdahl and the convex structure of reduced
  density matrices.
\newblock \emph{Journal of Mathematical Physics}, 53\penalty0 (7):\penalty0
  072203, 2012{\natexlab{a}}.

\bibitem[Chen et~al.(2012{\natexlab{b}})Chen, Ji, Zeng, and
  Zhou]{chen2012ground}
Jianxin Chen, Zhengfeng Ji, Bei Zeng, and DL~Zhou.
\newblock From ground states to local hamiltonians.
\newblock \emph{Physical Review A}, 86\penalty0 (2):\penalty0 022339,
  2012{\natexlab{b}}.

\bibitem[Chen et~al.(2013)Chen, Dawkins, Ji, Johnston, Kribs, Shultz, and
  Zeng]{chen2013uniqueness}
Jianxin Chen, Hillary Dawkins, Zhengfeng Ji, Nathaniel Johnston, David Kribs,
  Frederic Shultz, and Bei Zeng.
\newblock Uniqueness of quantum states compatible with given measurement
  results.
\newblock \emph{Physical Review A}, 88\penalty0 (1):\penalty0 012109, 2013.

\bibitem[Qi et~al.(2013)Qi, Hou, Li, Dong, Xiang, and Guo]{qi2013quantum}
Bo~Qi, Zhibo Hou, Li~Li, Daoyi Dong, Guoyong Xiang, and Guangcan Guo.
\newblock Quantum state tomography via linear regression estimation.
\newblock \emph{Scientific reports}, 3:\penalty0 3496, 2013.

\bibitem[Zeng et~al.(2015)Zeng, Chen, Zhou, and Wen]{zeng2015quantum}
Bei Zeng, Xie Chen, Duan-Lu Zhou, and Xiao-Gang Wen.
\newblock Quantum information meets quantum matter--from quantum entanglement
  to topological phase in many-body systems.
\newblock \emph{arXiv preprint arXiv:1508.02595}, 2015.

\bibitem[Kieferov{\'a} and Wiebe(2017)]{kieferova2017tomography}
M{\'a}ria Kieferov{\'a} and Nathan Wiebe.
\newblock Tomography and generative training with quantum boltzmann machines.
\newblock \emph{Physical Review A}, 96\penalty0 (6):\penalty0 062327, 2017.

\bibitem[Torlai et~al.(2018)Torlai, Mazzola, Carrasquilla, Troyer, Melko, and
  Carleo]{torlai2018neural}
Giacomo Torlai, Guglielmo Mazzola, Juan Carrasquilla, Matthias Troyer, Roger
  Melko, and Giuseppe Carleo.
\newblock Neural-network quantum state tomography.
\newblock \emph{Nature Physics}, 14\penalty0 (5):\penalty0 447, 2018.
\newblock URL \url{https://www.nature.com/articles/s41567-018-0048-5}.

\bibitem[Le~Roux and Bengio(2008)]{le2008representational}
Nicolas Le~Roux and Yoshua Bengio.
\newblock Representational power of restricted boltzmann machines and deep
  belief networks.
\newblock \emph{Neural computation}, 20\penalty0 (6):\penalty0 1631--1649,
  2008.

\bibitem[Kingma and Ba(2014)]{kingma2014adam}
Diederik~P Kingma and Jimmy Ba.
\newblock Adam: A method for stochastic optimization.
\newblock \emph{arXiv preprint arXiv:1412.6980}, 2014.

\bibitem[Reddi et~al.(2018)Reddi, Kale, and Kumar]{reddi2018convergence}
Sashank~J Reddi, Satyen Kale, and Sanjiv Kumar.
\newblock On the convergence of adam and beyond.
\newblock 2018.

\bibitem[Qi and Ranard(2017)]{qi2017determining}
Xiao-Liang Qi and Daniel Ranard.
\newblock Determining a local hamiltonian from a single eigenstate.
\newblock \emph{arXiv preprint arXiv:1712.01850}, 2017.

\bibitem[Nielsen and Chuang(2002)]{nielsen2002quantum}
Michael~A Nielsen and Isaac Chuang.
\newblock Quantum computation and quantum information, 2002.

\bibitem[Xin et~al.(2018{\natexlab{a}})Xin, Wang, Li, Kong, Wei, Wang, Ruan,
  and Long]{xin2018nuclear}
Tao Xin, Bi-Xue Wang, Ke-Ren Li, Xiang-Yu Kong, Shi-Jie Wei, Tao Wang, Dong
  Ruan, and Gui-Lu Long.
\newblock Nuclear magnetic resonance for quantum computing: Techniques and
  recent achievements.
\newblock \emph{Chinese Physics B}, 27\penalty0 (2):\penalty0 020308,
  2018{\natexlab{a}}.

\bibitem[Vandersypen and Chuang(2005)]{vandersypen2005nmr}
Lieven~MK Vandersypen and Isaac~L Chuang.
\newblock Nmr techniques for quantum control and computation.
\newblock \emph{Reviews of modern physics}, 76\penalty0 (4):\penalty0 1037,
  2005.

\bibitem[Jones et~al.(2000)Jones, Vedral, Ekert, and
  Castagnoli]{jones2000geometric}
Jonathan~A Jones, Vlatko Vedral, Artur Ekert, and Giuseppe Castagnoli.
\newblock Geometric quantum computation using nuclear magnetic resonance.
\newblock \emph{Nature}, 403\penalty0 (6772):\penalty0 869, 2000.

\bibitem[Xin et~al.(2018{\natexlab{b}})Xin, Huang, Lu, Li, Luo, Yin, Li, Lu,
  Long, and Zeng]{xin2018nmrcloudq}
Tao Xin, Shilin Huang, Sirui Lu, Keren Li, Zhihuang Luo, Zhangqi Yin, Jun Li,
  Dawei Lu, Guilu Long, and Bei Zeng.
\newblock Nmrcloudq: a quantum cloud experience on a nuclear magnetic resonance
  quantum computer.
\newblock \emph{Science Bulletin}, 63\penalty0 (1):\penalty0 17--23,
  2018{\natexlab{b}}.

\bibitem[Cory et~al.(1997)Cory, Fahmy, and Havel]{cory1997ensemble}
David~G Cory, Amr~F Fahmy, and Timothy~F Havel.
\newblock Ensemble quantum computing by nmr spectroscopy.
\newblock \emph{Proceedings of the National Academy of Sciences}, 94\penalty0
  (5):\penalty0 1634--1639, 1997.

\bibitem[Fahmy and Havel(2000)]{fahmy2000nuclear}
Amr~F Fahmy and Timothy~F Havel.
\newblock Nuclear magnetic resonance spectroscopy: An experimentally accessible
  paradigm for quantum computing.
\newblock \emph{Quantum Computation and Quantum Information Theory: Reprint
  Volume with Introductory Notes for ISI TMR Network School, 12-23 July 1999,
  Villa Gualino, Torino, Italy}, page 471, 2000.

\bibitem[Knill et~al.(1998)Knill, Chuang, and Laflamme]{knill1998effective}
Emanuel Knill, Isaac Chuang, and Raymond Laflamme.
\newblock Effective pure states for bulk quantum computation.
\newblock \emph{Physical Review A}, 57\penalty0 (5):\penalty0 3348, 1998.

\bibitem[Schmidhuber(2015)]{schmidhuber2015deep}
J{\"u}rgen Schmidhuber.
\newblock Deep learning in neural networks: An overview.
\newblock \emph{Neural networks}, 61:\penalty0 85--117, 2015.

\bibitem[Chollet et~al.(2015)]{chollet2015keras}
Fran\c{c}ois Chollet et~al.
\newblock Keras.
\newblock \url{https://keras.io}, 2015.

\bibitem[Abadi et~al.(2016)Abadi, Barham, Chen, Chen, Davis, Dean, Devin,
  Ghemawat, Irving, Isard, et~al.]{abadi2016tensorflow}
Mart{\'\i}n Abadi, Paul Barham, Jianmin Chen, Zhifeng Chen, Andy Davis, Jeffrey
  Dean, Matthieu Devin, Sanjay Ghemawat, Geoffrey Irving, Michael Isard, et~al.
\newblock Tensorflow: a system for large-scale machine learning.
\newblock In \emph{OSDI}, volume~16, pages 265--283, 2016.

\bibitem[Nair and Hinton(2010)]{nair2010rectified}
Vinod Nair and Geoffrey~E Hinton.
\newblock Rectified linear units improve restricted boltzmann machines.
\newblock In \emph{Proceedings of the 27th international conference on machine
  learning (ICML-10)}, pages 807--814, 2010.

\bibitem[Gershenfeld and Chuang(1997)]{gershenfeld1997bulk}
Neil~A Gershenfeld and Isaac~L Chuang.
\newblock Bulk spin-resonance quantum computation.
\newblock \emph{science}, 275\penalty0 (5298):\penalty0 350--356, 1997.

\bibitem[Boulant et~al.(2003)Boulant, Edmonds, Yang, Pravia, and
  Cory]{boulant2003experimental}
N~Boulant, K~Edmonds, J~Yang, MA~Pravia, and DG~Cory.
\newblock Experimental demonstration of an entanglement swapping operation and
  improved control in nmr quantum-information processing.
\newblock \emph{Physical Review A}, 68\penalty0 (3):\penalty0 032305, 2003.

\bibitem[Khaneja et~al.(2005)Khaneja, Reiss, Kehlet, Schulte-Herbr{\"u}ggen,
  and Glaser]{khaneja2005optimal}
Navin Khaneja, Timo Reiss, Cindie Kehlet, Thomas Schulte-Herbr{\"u}ggen, and
  Steffen~J Glaser.
\newblock Optimal control of coupled spin dynamics: design of nmr pulse
  sequences by gradient ascent algorithms.
\newblock \emph{Journal of magnetic resonance}, 172\penalty0 (2):\penalty0
  296--305, 2005.

\bibitem[Ryan et~al.(2008)Ryan, Negrevergne, Laforest, Knill, and
  Laflamme]{ryan2008liquid}
CA~Ryan, C~Negrevergne, M~Laforest, E~Knill, and R~Laflamme.
\newblock Liquid-state nuclear magnetic resonance as a testbed for developing
  quantum control methods.
\newblock \emph{Physical Review A}, 78\penalty0 (1):\penalty0 012328, 2008.

\bibitem[Lu et~al.(2017)Lu, Li, Li, Katiyar, Park, Feng, Xin, Li, Long,
  Brodutch, et~al.]{lu2017enhancing}
Dawei Lu, Keren Li, Jun Li, Hemant Katiyar, Annie~Jihyun Park, Guanru Feng, Tao
  Xin, Hang Li, Guilu Long, Aharon Brodutch, et~al.
\newblock Enhancing quantum control by bootstrapping a quantum processor of 12
  qubits.
\newblock \emph{npj Quantum Information}, 3\penalty0 (1):\penalty0 45, 2017.

\bibitem[Leskowitz and Mueller(2004)]{leskowitz2004state}
Garett~M Leskowitz and Leonard~J Mueller.
\newblock State interrogation in nuclear magnetic resonance quantum-information
  processing.
\newblock \emph{Physical Review A}, 69\penalty0 (5):\penalty0 052302, 2004.

\bibitem[Lee(2002)]{lee2002quantum}
Jae-Seung Lee.
\newblock The quantum state tomography on an nmr system.
\newblock \emph{Physics Letters A}, 305\penalty0 (6):\penalty0 349--353, 2002.

\bibitem[Li et~al.(2017{\natexlab{b}})Li, Huang, Luo, Li, Lu, and
  Zeng]{PhysRevA.96.032307}
Jun Li, Shilin Huang, Zhihuang Luo, Keren Li, Dawei Lu, and Bei Zeng.
\newblock Optimal design of measurement settings for quantum-state-tomography
  experiments.
\newblock \emph{Phys. Rev. A}, 96:\penalty0 032307, Sep 2017{\natexlab{b}}.
\newblock \doi{10.1103/PhysRevA.96.032307}.
\newblock URL \url{https://link.aps.org/doi/10.1103/PhysRevA.96.032307}.

\bibitem[Altepeter et~al.(2005)Altepeter, Jeffrey, and
  Kwiat]{altepeter2005photonic}
Joseph~B Altepeter, Evan~R Jeffrey, and Paul~G Kwiat.
\newblock Photonic state tomography.
\newblock \emph{Advances in Atomic, Molecular, and Optical Physics},
  52:\penalty0 105--159, 2005.

\bibitem[Song et~al.(2017)Song, Xu, Liu, Yang, Zheng, Deng, Xie, Huang, Guo,
  Zhang, Zhang, Xu, Zheng, Zhu, Wang, Chen, Lu, Han, and
  Pan]{PhysRevLett.119.180511}
Chao Song, Kai Xu, Wuxin Liu, Chui-ping Yang, Shi-Biao Zheng, Hui Deng, Qiwei
  Xie, Keqiang Huang, Qiujiang Guo, Libo Zhang, Pengfei Zhang, Da~Xu, Dongning
  Zheng, Xiaobo Zhu, H.~Wang, Y.-A. Chen, C.-Y. Lu, Siyuan Han, and Jian-Wei
  Pan.
\newblock 10-qubit entanglement and parallel logic operations with a
  superconducting circuit.
\newblock \emph{Phys. Rev. Lett.}, 119:\penalty0 180511, Nov 2017.
\newblock \doi{10.1103/PhysRevLett.119.180511}.
\newblock URL \url{https://link.aps.org/doi/10.1103/PhysRevLett.119.180511}.

\bibitem[Krizhevsky et~al.(2012)Krizhevsky, Sutskever, and
  Hinton]{krizhevsky2012imagenet}
Alex Krizhevsky, Ilya Sutskever, and Geoffrey~E Hinton.
\newblock Imagenet classification with deep convolutional neural networks.
\newblock In \emph{Advances in neural information processing systems}, pages
  1097--1105, 2012.

\bibitem[He et~al.(2016)He, Zhang, Ren, and Sun]{he2016deep}
Kaiming He, Xiangyu Zhang, Shaoqing Ren, and Jian Sun.
\newblock Deep residual learning for image recognition.
\newblock In \emph{Proceedings of the IEEE conference on computer vision and
  pattern recognition}, pages 770--778, 2016.

\end{thebibliography}

\end{document}